\documentclass{PoS}

\newcommand{\bea}{\begin{eqnarray}}
\newcommand{\eea}{\end{eqnarray}}
\newcommand{\oh}{\frac{1}{2}}

\title{Phase diagram of the three dimensional Thirring model - A Monte Carlo study}

\ShortTitle{Three dimensional Thirring model}

\author{Stavros Christofi \\ 
        Frederick Institute of Technology, CY-1303 Nicosia, Cyprus \\ 
        \email{eng.cs@fit.ac.cy}} 
\author{Simon Hands \\
        Physics Department, Swansea University, Singleton Park, Swansea SA2 8PP, U.K. \\
        \email{s.hands@swan.ac.uk}}
\author{\speaker{Costas Strouthos} \\
        Harvard-MIT (HST) Martinos Center for Biomedical Imaging, 
        Massachusetts General Hospital, Harvard Medical School, 
        Charlestown, MA 02129, U.S.A. \\
        \email{cstrouth@deas.harvard.edu}}

\abstract{
 Certain approximate solutions of the continuum Schwinger-Dyson Equations (SDEs)
 predict chiral symmetry breaking in the 3d Thirring model
 when the number of fermion flavors $N_f<4.32$ \cite{itoh} whereas others predict symmetry 
 breaking for all $N_f$ \cite{hong}. 
 Our results from Monte Carlo simulations with $N_f=6$, predict a second order chiral phase transition.
 The critical coupling in this case corresponds to an ultra-violet fixed point of the
 renormalization group defining a non-trivial continuum limit.
 Further, our numerical simulations provide an estimate 
 for the critical number of fermion flavors, $N_{fc} \approx 6.5$.
}

\FullConference{XXIVth International Symposium on Lattice Field Theory\\
                July 23-28, 2006\\
                Tucson, Arizona, USA}

\begin{document}

\section{Introduction}
The study of quantum field theories in which the ground state shows a sensitivity to the number of fermion 
flavors $N_f$ is intrinsically interesting. According to certain approximate solutions of continuum SDEs 
the 3d Thirring model as well as QED$_3$ display this phenomenon.
The Thirring model is a theory of fermions interacting via a current contact interaction:
\begin{equation}
{\cal L}=\bar\psi_i(\partial{\!\!\!\!\! /}\,+m)\psi_i
+{g^2\over{2N_f}}(\bar\psi_i\gamma_\mu\psi_i)^2,
\end{equation}
where $\psi_i,\bar\psi_i$ are four-component spinors, $m$ is a 
parity-conserving bare mass, and the index $i$ runs over $N_f$ distinct
fermion flavors. 
In the chiral limit the Lagrangian of the model has a continuous $U(1)$ chiral symmetry.
Since the coupling $g^2$ has mass dimension $-1$, naive
power-counting suggests that the model is non-renormalisable.
However, as has been suspected for many years \cite{parisi,gomes,hands1}, an
expansion in powers of $1/N_f$, rather than $g^2$, is exactly renormalisable
and the model has a well-defined continuum limit corresponding to a UV-stable
fixed point of the renormalization group.
After the introduction of an
auxiliary vector field, the above Lagrangian can be rewritten as:
\begin{equation}
        {\cal L}_{aux} = \bar\psi_i(\partial{\!\!\!\!\! /}\,
        + iA{\!\!\!\!\! /}\, + m)\psi_i
        + {N_f\over{2g^2}} (A_\mu)^2.
\label{eq:auxiliary}
\end{equation}
The $1/N_f$ expansion may not, however, describe the true behavior of the model,
particularly for small $N_f$. Chiral symmetry breaking is forbidden at all orders
in $1/N_f$, and yet may be predicted by a self-consistent approach such as
SDEs \cite{itoh,hong}. SDEs solved exactly in the limit $g^2\rightarrow \infty$ \cite{itoh} 
show that chiral symmetry is broken for $N_f < N_{fc} \simeq 4.32$.
This is close to certain predictions of $N_{fc}$  for the non-trivial IR behavior in QED$_3$ \cite{qed3}.
Based on these results, at $N_{fc}$ the model is expected to undergo an infinite order or conformal
phase transition, originally discussed by Miranskii {\it et al} in the context of quenched QED$_4$ \cite{miranskii}.
Using a different sequence of truncations Hong and Park \cite{hong} found chiral symmetry breaking for all $N_f$.
Previous Monte Carlo simulations \cite{hands2,hands3,hands4,hands5} provided evidence that for $N_f \leq 5$
the model is in the chirally broken phase at strong enough coupling.
Here, we present preliminary results based on numerical simulations with $N_f=2,...,18$ 
in an effort to estimate $N_{fc}$.

\section{Lattice Model}
The lattice action we have used is based on \cite{hands2} and is as follows:
\begin{eqnarray}
      S &=& \oh \sum_{x\mu i} \bar\chi_i(x) \eta_\mu(x)
        (1+iA_\mu(x)) \chi_i(x+\hat\mu)
        + \mbox{h.c.} \nonumber \\
          & & + m \sum_{xi} \bar\chi_i(x) \chi_i(x) +
        \frac{N}{4g^2} \sum_{x\mu} A_\mu(x)^2 \nonumber \\
          &\equiv& \sum_{xyi}\bar\chi_i(x) M[A,m](x,y) \chi_i(y) +
          \frac{N}{4g^2} \sum_{x\mu} A_\mu(x)^2,
\label{eq:lat}
\end{eqnarray}
where $\chi,\bar\chi$ are staggered fermion fields and the flavor index $i$ runs
from 1 to $N$. We have introduced $M[A,m]$ for the fermionic bilinear, which
depends on both the auxiliary field defined on the lattice links and the bare mass $m$.
In three dimensions $N$ staggered fermion describe $N_f=2N$ four-component continuum species. 
We employed the Hybrid Monte Carlo (HMC) algorithm to simulate even $N_f$ and  
we used the Hybrid Molecular Dynamics (HMD) algorithm for odd or non-integer $N_f$. 
In the HMD simulations we used small enough fictitious time step $\Delta \tau \leq 0.0025$
and ensured that the $O(N^2 \Delta \tau^2)$ systematic errors in the molecular dynamics steps 
were smaller than the statistical errors of the various observables. 
\begin{figure}
\begin{center}
\includegraphics[width=.5\textwidth]{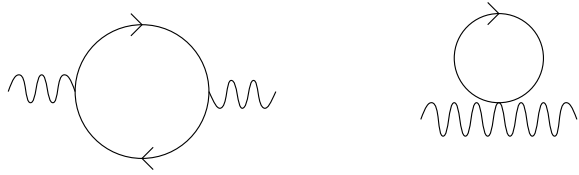}
\caption{Diagrams contributing to vacuum polarization in lattice QED.}
\label{fig:fig1a}
\end{center}
\end{figure}
\section{Critical Number of Fermion Flavors}
In this section, we summarize the discussion presented in \cite{hands3} regarding the non-conservation of the 
vector current at strong coupling in the lattice Thirring model of eq.~(\ref{eq:lat}). 
At leading order in $1/N$ the vector propagator receives an extra
non-transverse contribution from vacuum polarization, essentially due to the absence of the second diagram
in Fig.~\ref{fig:fig1a}. 
It should be noted that in non-compact QED$_3$ both diagrams are present and
a cancellation of O($a^{-1}$) contributions from the two diagrams in Fig.~\ref{fig:fig1a} occurs.
In the 3d Thirring model however, the effect of the absence of the second diagram
can be absorbed into a redefinition of the coupling:
\begin{equation}
g_R^2={g^2\over{1-g^2J(m)}},
\end{equation}
where $J(m)$ is the value of the integral contributed from the second diagram in the above figure.
\begin{figure}
\begin{center}
\includegraphics[width=.6\textwidth]{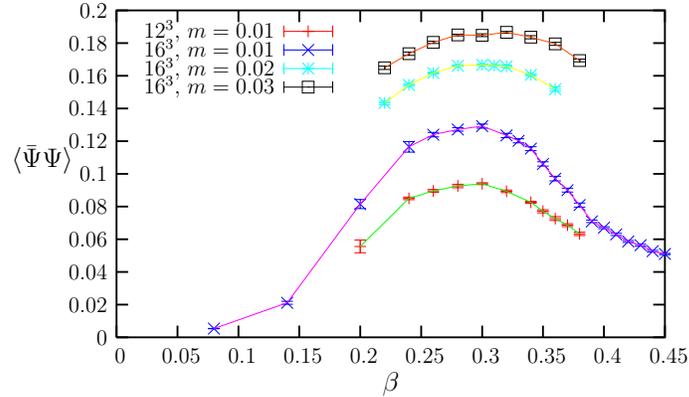}
\caption{Chiral condensate vs. $\beta$ for $N_f=6$ near $\beta_{\rm lim}$.} 
\label{fig:fig2}
\end{center}
\end{figure}
\begin{figure}
\begin{center}
\includegraphics[width=.6\textwidth]{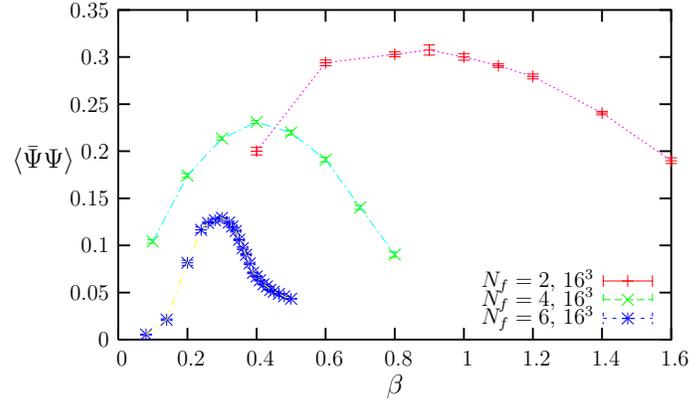}
\caption{Chiral condensate vs. $\beta$ for $N_f=2,4,6$ on $16^3$ lattices with $m=0.01$.}
\label{fig:fig3}
\end{center}
\end{figure}
\begin{figure}
\begin{center}
\includegraphics[width=.6\textwidth]{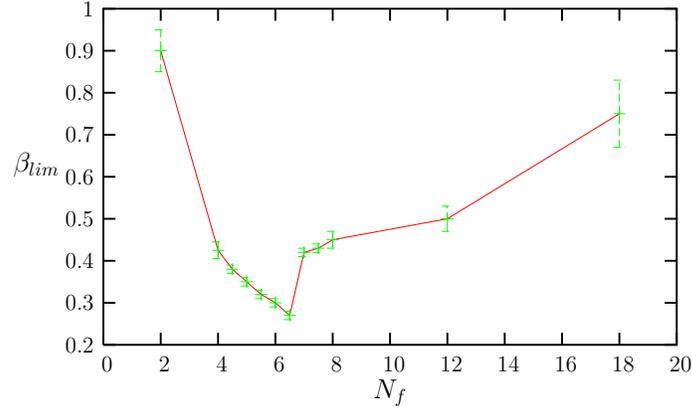}
\caption{$\beta_{\rm lim}$ vs. $N_f$.}
\label{fig:fig4}
\end{center}
\end{figure}
The physics described by continuum $1/N_f$ perturbation theory occurs for
the range of couplings $g_R^2\in[0,\infty)$, ie. for $g^2\in[0,g_{\rm lim}^2)$;
to leading order in $1/N$ ${1/g_{\rm lim}^2}={2/3}$ for $m=0$.
In summary, the strong coupling limit $1/g^2 \rightarrow 0$ of the continuum model
is recovered in the lattice model at $1/g^2 \rightarrow g_{\rm lim}^2 > 0$.
As we already discussed in the Introduction, chiral symmetry breaking is absent in large-N calculations.
Therefore, it may well be that the value of the second diagram in Fig.~\ref{fig:fig1a} is considerably altered 
in a chirally broken vacuum. Later in this section we will use this strong coupling 
lattice discretization effect as a criterion to decide
whether chiral symmetry is broken or not at infinite coupling for different values of $N_f$.
The data presented in  Fig.~\ref{fig:fig2} show clearly that the value of $\beta_{\rm lim}$ (which corresponds to the 
infinite coupling limit in the continuum), where the condensate has its maximum value does not depend on the 
lattice size, because as mentioned in an earlier paragraph this is an $O(a^{-1})$ discretization effect. In addition, 
the same figure shows that within our statistical errors the position of the peak does not depend on the lattice
size or the bare mass.
Figure \ref{fig:fig3} shows that as expected the value of the chiral condensate (for $m=0.01$) decreases as $N_f$ is increased, 
because the fermion-antifermion
screening has a stronger contribution for larger $N_f$. Furthermore, the value of $\beta_{\rm lim}$ shifts to the left 
as the number of fermion flavors increases from $N_f=2$ to $N_f=6$. 
Figure \ref{fig:fig4} shows that this trend is not monotonic; near $N_f \approx 6.5$, the curve of $\beta_{\rm lim}(N_f)$ 
reaches a minimum, implying a significant change in the strong coupling behavior of the model. 
The minimum in $\beta_{\rm lim}(N_f)$ is a significant change 
in the phase diagram of the 3d Thirring model,
implying that $N_{fc} \approx 6.5$. This conclusion is enhanced by the results shown in Fig.~\ref{fig:fig5}, 
where we plot the chiral condensate at $\beta_{\rm lim}(N_f)$ versus $N_f$. The relatively rapid change 
in the value of the condensate around $N_f=6.5$ may imply that a phase transition separates a chirally broken 
from a chirally symmetric phase with $N_{fc} \approx 6.5$. We are currently extending our simulations 
in order to understand better the properies of the chiral symmetry 
restoration transition.
\begin{figure}
\begin{center}
\includegraphics[width=.6\textwidth]{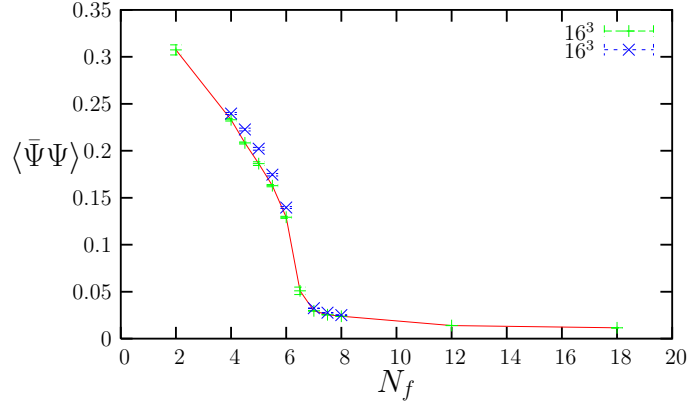}
\caption{Chiral condensate at $\beta_{\rm lim}$ for different values of $N_f$.}
\label{fig:fig5}
\end{center}
\end{figure}
\section{Analysis of $N_f=6$ data}
\begin{figure}
\begin{center}
\includegraphics[width=.58\textwidth]{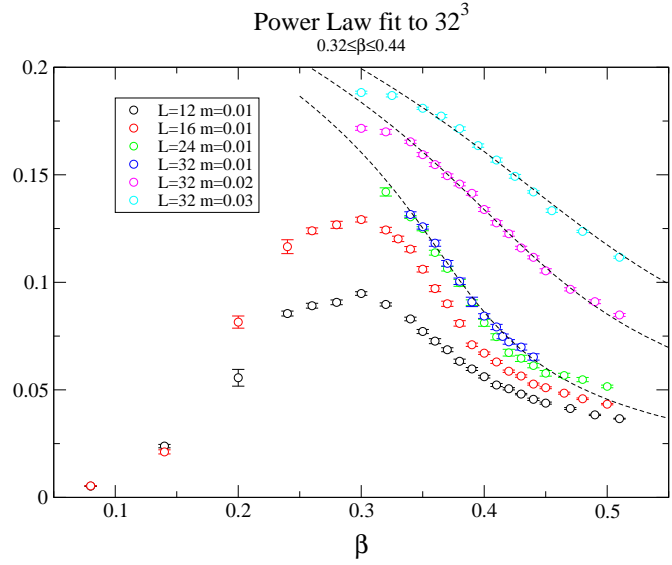}
\caption{Chiral condensate vs. $\beta$ for $N_f=6$. The dashed lines represent 
fits of the $32^3$, $0.32 \leq \beta \leq 0.44$ data to eq.~4.1.}
\label{fig:fig6}
\end{center}
\end{figure}

During the last decade, Monte Carlo simulations provided evidence that for $N_f \leq 5$ and for 
sufficiently strong coupling the 3d Thirring model is in the chirally broken phase \cite{hands2,hands3,hands4,hands5}. 
In this section we present results from new simulations with $N_f=6$ on lattice sizes varying from $12^3$ to $32^3$. 
As expected the phase transition for $N_f=6$ is steeper than the transitions
with smaller $N_f$. However, a detailed finite size scaling analysis of the Binder's Cumulant (not shown here)
showed that the data at $m=0.01$ are consistent with a crossover instead of a first order transition.
In order to understand the properties of the model near the transition we fitted the data for the chiral condensate extracted 
from simulations on the largest $32^3$ lattices to
the Equation of State (EoS):
\begin{equation}
m=A(\beta-\beta_c)\langle\bar\chi\chi\rangle^p+
B\langle\bar\chi\chi\rangle^\delta;
\label{eq:eos1}
\end{equation}
where $p=\delta-1/\beta_m$.
The assumption in eq.~(\ref{eq:eos1}) that finite size effects are negligible is justified by the
observation that for the smallest mass used, $m=0.01$, the difference between
the values of $\langle \bar{\Psi} \Psi\rangle$ 
extracted from simulations on $24^3$ and $32^3$ lattices is small.
The fit was performed for $\beta = 0.34,...,0.44$ (see fig.~\ref{fig:fig6}), i.e. 
close enough to the transition but still away from
$\beta_{lim} \approx 0.30$ where strong coupling lattice discretization effects are severe.
The results are: $A=1.38(13)$, $B=58(25)$, $\beta_c=0.3082(36)$, $\delta=4.65(27)$, and $\beta_m=0.28(2)$
with an acceptable fit quality $\chi^2/{\rm dof}=1.6$.
For $N_f=6$ the value $\beta_c$ is close to
$\beta_{lim}$, implying that $N_f=6$ is close to $N_{fc}$, which is in accordance with the results
presented in section 4.
We also performed fits which included data from smaller volumes to a finite size scaling 
form of the EoS: 
\begin{equation}
m_0=A((\beta-\beta_c)+CL^{-{1\over\nu}})\langle\bar\chi\chi\rangle^p+
B\langle\bar\chi\chi\rangle^\delta
\label{eq:eos2}
\end{equation}
\begin{figure}
\begin{center}
\includegraphics[width=.58\textwidth]{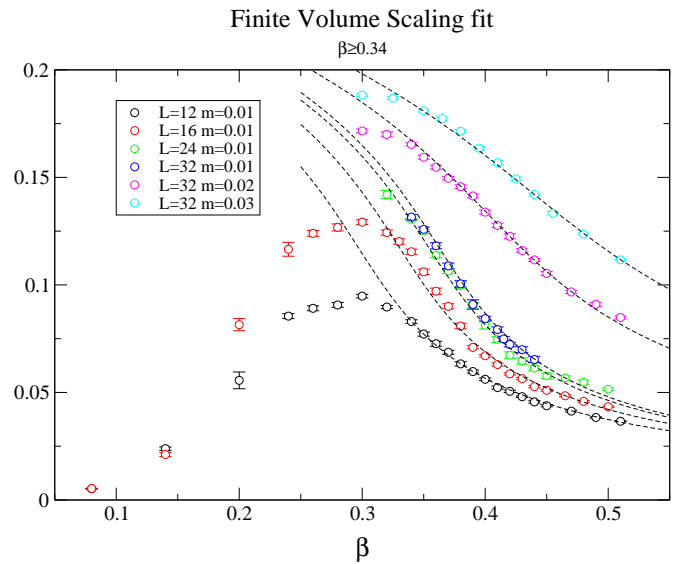}
\caption{Chiral condensate vs. $\beta$ for $N_f=6$. The dashed lines represent fits of data 
from all lattice sizes for $\beta \geq 0.34$ to eq.~4.2.} 
\label{fig:fig7}
\end{center}
\end{figure}
The results extracted from fits to eq.~(\ref{eq:eos2}) (see fig.~\ref{fig:fig7}) are of lower quality ($\chi^2/dof=5.8$) 
because on smaller volumes the peaks near $\beta_{lim}$ are broader, implying that strong coupling lattice discretization
effects are more severe. 
In this case we got $A=1.91(4)$, $B=231(53)$,
$\beta=0.3185(10)$, $\delta=5.49(13)$, and $\beta_m=0.19(2)$.

\section{Conclusions}
Our main results is that the $N_f=6$ 3d Thirring model like its smaller $N_f$ counterparts appears to 
undergo a second order chiral symmetry breaking phase transition at strong coupling. Moreover, the $N_f=6$ critical exponents
are distinct from those of $N_f=2$ and $N_f=4$ i.e. the different models have interacting continuum limits which are
qualitatively, but not quantitatively similar. Further, a detailed analysis of the strong coupling behavior of the model 
for $N_f=2,...,18$ revealed that the critical number of fermion flavors where chiral symmetry is restored is 
$N_{fc} \approx 6.5$ which is distinctly larger than the $N_{fc}=4.32$ predicted by certain SDE approaches \cite{itoh}
and it contradicts the conclusion drawn from other SDE approaches \cite{hong} which predict chiral symmetry breaking for all 
values of $N_f$. Our results may have broader interesting implications on the applicability of SDEs in 
non-perturbative field theories. There are predictions \cite{parisi,gomes,hands1} that in the strong coupling limit 
the vector meson becomes massless suggesting that the IR limit of QED$_3$ coincides with the UV limit of the 
3d Thirring model. The problem of detecting dynamical symmetry breaking in lattice QED$_3$ is much harder 
than in 3d Thirring, because in QED$_3$ the dynamical fermion mass is much smaller than the momentum cut-off \cite{qed3b}.

\section*{Acknowledgements}
\noindent
The simulations were performed on a cluster of 64-bit AMD Opterons 250 at the Frederick Institute of 
Technology, Cyprus.



\begin{thebibliography}{99}
\bibitem{itoh} T. Itoh, Y. Kim, M. Sugiura and K. Yamawaki, Prog. Theor. Phys. {\bf 93} (1995) 417 [{\tt hep-th/9411201}]. \\
\bibitem{hong} D.K. Hong and S.H. Park, Phys. Rev. {\bf D49} (1994) 5507 [{\tt hep-th/9307186}]. \\
\bibitem{parisi} G. Parisi, Nucl. Phys. {\bf B100} (1975) 368; 
S. Hikami and T. Muta, Prog. Theor. Phys. {\bf 57} (1977) 785;  
Z. Yang, Texas preprint UTTG-40-90 (1990). \\
\bibitem{gomes} M. Gomes, R.S. Mendes, R.F. Ribeiro and A.J. da Silva, Phys. Rev.
{\bf D43} (1991) 3516. \\
\bibitem{hands1} S.J. Hands, Phys. Rev. {\bf D51} (1995) 5816 [{\tt hep-th/9411016}]. \\
\bibitem{qed3}T. Appelquist, M. Bowick, D. Karabali and L.C.R.
Wijewardhana, Phys. Rev. {\bf D33} (1986) 3704, 3774;
M.R. Pennington and D. Walsh, Phys. Lett. {\bf B253} (1991) 246;
P. Maris, Phys. Rev. {\bf D54} (1996) 4049 [{\tt hep-ph/9406214}].
\bibitem{miranskii} V.A. Miranskii, Nuovo Cimento {\bf 90A} (1985) 149. \\
\bibitem{hands2} L. Del Debbio and S. Hands, Phys. Lett {\bf B373} (1996) 171 [{\tt hep-lat/9512013}].  \\ 
\bibitem{hands3} L. Del Debbio, S. Hands, and J.C. Mehegan, Nucl. Phys. {\bf B502} (1997) 269 [{\tt hep-lat/9701016}]. \\ 
\bibitem{hands4} L. Del Debbio and S. Hands, Nucl. Phys. {\bf B552} (1999) 339 [{\tt hep-lat/9902014}]. \\
\bibitem{hands5} S. Hands and B. Lucini, Phys. Lett. {\bf B461} (1999) 263 [{\tt hep-lat/9906008}]. \\
\bibitem{qed3b} S. Hands, J. Kogut, C. Strouthos, Nucl. Phys. {\bf B645} (2002) 321 [{\tt hep-lat/0208030}]; 
S. Hands, J. Kogut, C. Strouthos, Phys. Rev. {\bf B70} (2004) 104501 [{\tt hep-lat/0404013}].

\end{thebibliography}
\end{document}